\documentclass{article}
\usepackage{psfig,spie,xspace,aasjournals}
 
\def\uchii{{UCH\textsc{ii}}\xspace}

\def\arcsec{\hbox{$^{\prime\prime}$}\xspace} \def\brg{\rm
  Br\ensuremath{\gamma}\xspace}
\def\farcs{\hbox{$.\!\!^{\prime\prime}$}}

 \def\msun{\rm
  M\ensuremath{_{\odot}}\xspace}
   
 \def\cmc{\rm cm\ensuremath{^{-3}}\xspace}

\def\mum{\ensuremath{\mu}m\xspace}
\newcommand{\ttt}[1]{\ensuremath{\times 10^{#1}}}
\title{The Impact of Adaptive Optics on Star Formation Research}

\author{M. Feldt\supit{a}, M. Kasper\supit{a}, F. Eisenhauer\supit{b}
  and S. Hippler\supit{a}\skiplinehalf
  \supit{a}Max-Planck-Institut f\"ur Astronomie, \\ Heidelberg, Bundesrepublik Deutschland\\
  \supit{b}Max-Planck-Institut f\"ur Extraterrestrische Physik, \\
  Garching, Bundesrepublik Deutschland }

\authorinfo{Send correspondence to M.  Feldt, mfeldt@mpia-hd.mpg.de}

\begin{document} 
  \maketitle

\begin{abstract}
  In this paper, we discuss the benefits of ground-based, adaptive
  optics (AO) aided observations for star formation research.  After
  outlining the general advantages, we present results obtained during
  the ALFA science demonstration programme in 1999. These results
  underline the absolute necessity of AO assistance for almost any
  kind of observations regarding star formation regions.
\end{abstract}



\section{INTRODUCTION}
\label{sect:intro}  
\begin{figure}
  \begin{center}
    \begin{tabular}{c}
      \psfig{figure=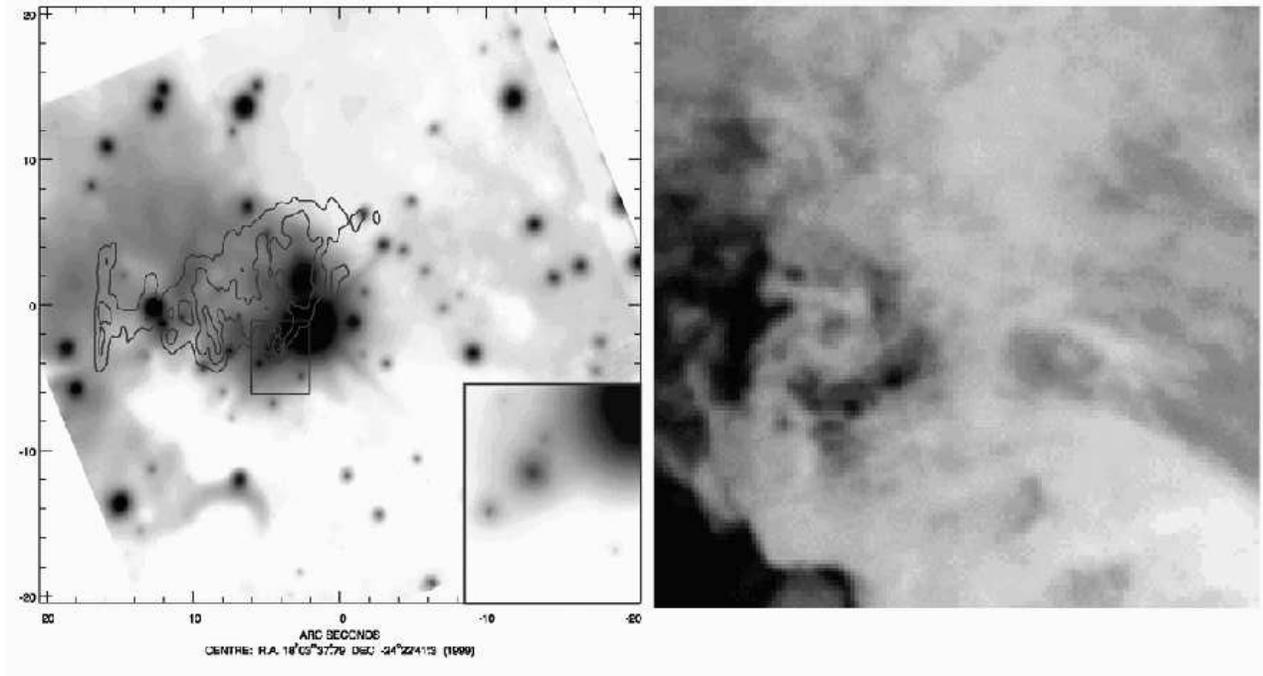,width=17cm,angle=270} 
    \end{tabular}
  \end{center}
  \caption[Her36] 
  { \label{fig:her36} The core of the Lagoon Nebula.  Left is an IR
    composite image taken with the AO system ALFA in the $J$,$H$, and
    $K'$-bands. On the right, you can see the same area observed with
    the HST\cite{stecklum:98}.  That is part of HST's famous
    ``Twisters in the Lagoon'' image taken with the WFPC2, do you
    recognise it? Note the number of obscured sources which are
    invisible in the HST image and the comparable resolution of the
    two frames.}
\end{figure}

Since about 1995, when adaptive optics (AO) was broadly introduced
into astronomical imaging techniques, the field of star formation has
seen revolutionary progress. Why is that? The first step towards the
new age had nothing to do with AO, it was the first direct imaging of
circumstellar disks provided by the HST (see, e.g McCaughrean et
al.\cite{mccaughrean:96}). Before the installation of the NICMOS
instrument, however, HST had the disadvantage of being restricted to
visual wavelengths.  Star formation, on the other hand, usually takes
place in molecular clouds. The radiation which carries information to
astronomers on earth must pass through these clouds. The dust
particles inside the clouds absorb most of the visual radiation, but
let the infrared (IR) part and longer wavelengths pass almost
untouched.  Additionally, dusty structures in the immediate
environment of young stellar objects (YSOs) provide IR excess emission
by absorbing shorter wavelengths and re-radiating the energy in the
IR. While this effectively shuts off observations of the act of star
formation in the visual (and thus by the old HST), it matches
excellently with the facilities provided by AO systems. Current
astronomical AO systems\footnote{At least those available to the
  general, civilian community}, as well as the very first ones 5 years
ago, usually use the visible part of the electromagnetic spectrum to
analyse the incoming wavefront, and provide the IR part to the science
instrument. This has several reasons: First of all, it is much easier
to achieve plain wavefronts at longer wavelengths: The longer the
wavelength, the smaller the number of required actuators and the
longer the atmospheric coherence timescale. On the other hand, using a
dichroic beam splitter is more efficient than a
``normal'' beam splitter.  AO systems with IR wavefront sensors
(ADONIS at ESO now has one) or systems providing AO in the visible
(like e.g. the starfire optical range), have to share photons between
the wavefront analyser and the science instrument. This always imposes
a further limitation on the limiting magnitudes of these systems.  All
these factors make AO systems ideal for observing star formation
regions, apart from the fact that on large telescopes they can easily
beat the resolution provided by HST. An example comparison of a
typical star forming region is given in Figure \ref{fig:her36}.

Despite the HST delivering the first direct images of circumstellar
disks, the detailed examination of such objects was performed later
with AO systems\cite{close:97}. AO supported IR
observations yielded information on the grain size distribution and
the temperature variations inside such disks. Polarimetric studies of
such disks have also been conducted\cite{close:99}. AO observations
have also contributed to other fields of star formation research, like
the search for brown dwarfs\cite{martin:99} (Important to fill the gap
between theories of star and planet formation), the identification of
initial mass functions in young clusters, the phenomena of massive
star formation\cite{stecklum:98s} and problems connected with binarity
or multiplicity of recently formed stellar systems.

In this paper, we will presented selected results obtained by a single
AO system, ALFA ({\bf A}daptive Optics with a {\bf L}aser {\bf f}or
{\bf A}stronomy\cite{hippler:98}, which is run by
the two Max Planck Institutes for Astronomy and extraterrestrial
physics in Heidelberg and Garching, Germany. We concentrate on fields
of star formation research which are examined at the two institutes.
The results we present were obtained during the ALFA science
demonstration run 1999, most of them in September of that year. The
intention is to give a short introduction to the fields we are working
in and to present current results obtained with the ALFA system. All
data presented are still under analysis and will be thoroughly
presented in forthcoming papers. To provide opportunities to judge the
``real life'' performance of the AO system, all characteristic data
(like Strehl ratios, FWHMs etc.) are given for the finally reduced
data, but without any applied post-reduction techniques like
deconvolution or PSF-fitting.


\section{Determining the IMF of NGC 6611}
The initial mass functions (IMFs) of young clusters contain a wealth
of hints towards the influence of the environment on the formation
mechanisms of stars. Particularly interesting is the interaction
between massive stars and their winds and ionising radiation, and
low-mass stars forming in the same molecular cloud. Since the general
scenario of low-mass star formation from a collapsing cloud core via a
disk/outflow scenario to a protostar is more or less
known\cite{shu:93,beckwith:96}, many parameters can be derived from
the mass distribution\cite{brandl:99}.
Among the questions in this context are: Does the slope of the IMF
vary on small scales? Do low-mass stars in a starburst event form
together with the more massive stars or at different times?  It might
also not be to daring to ask if low-mass stars form at all in an
environment of violent, young massive stars.

\subsection{Observations and Data Reduction} 
\begin{figure}
   \begin{center}
   \begin{tabular}{c}
   \psfig{figure=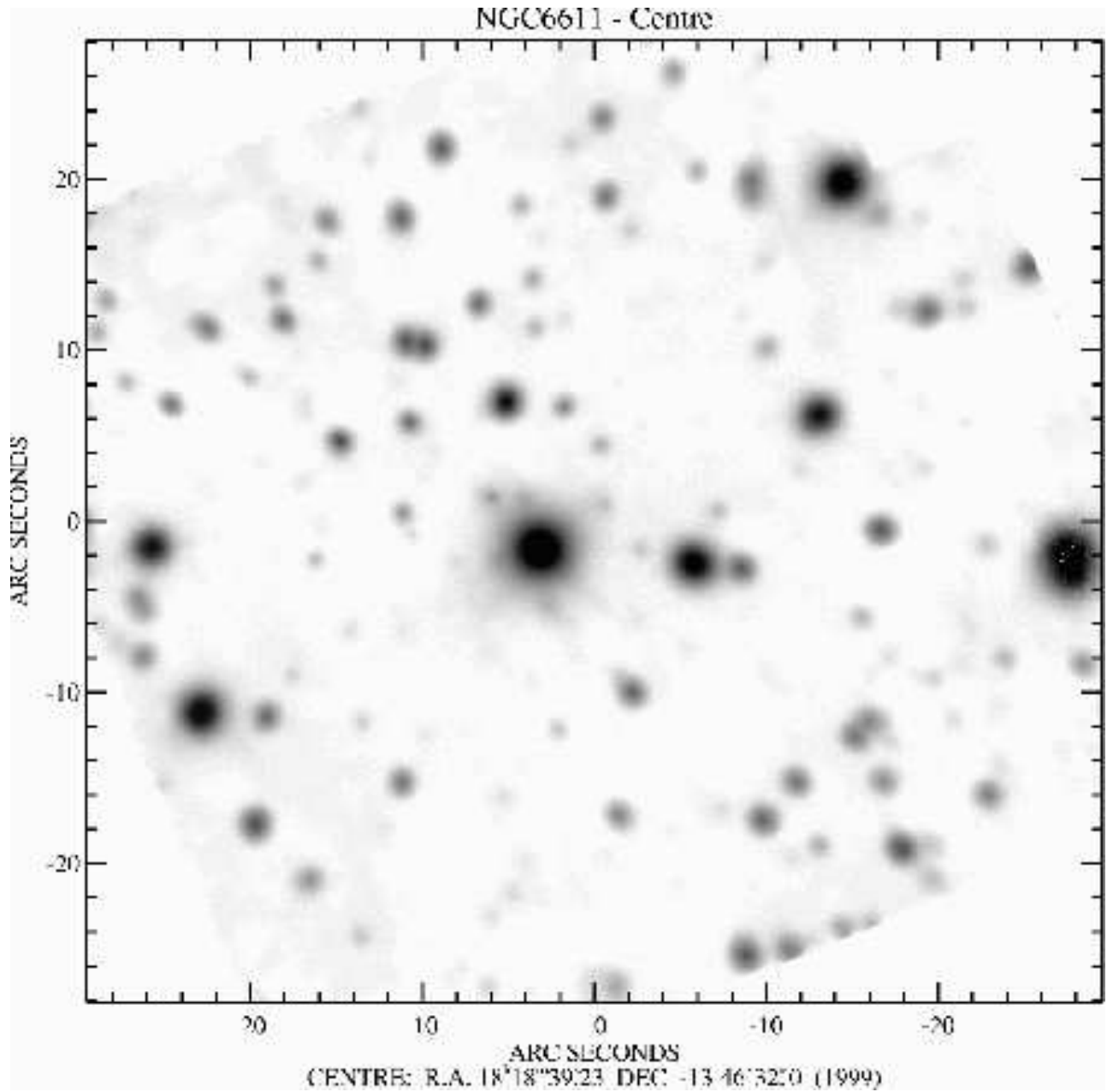,width=17cm} 
   \end{tabular}
   \end{center}
   \caption[ngc6611] 
   { \label{fig:ngc6611} The core of the young cluster NGC\,6611. The
     WFS is slightly left of the image centre, it has a brightness of
     9.0\,mag in $V$-band. Maximum Strehl in $K'$ is 8\%, the
     corresponding FWHM of the stellar PSF is 0\farcs28. All numbers
     are given for the final image, which consists of images
     registered to each other from 5 different mosaic positions. The
     effect of tip-tilt anisoplanatism is clearly visible, as stars
     towards the edges of the image appear elongated, the major axes
     of their ellipses pointing towards the WFS star.}
   \end{figure}

NGC\,6611 was observed during the ALFA science demonstration run in
September 1999. The seeing conditions were median with a $J$-band
seeing of 0\farcs9. Locked on the reference star of $m_V=9.0$ the AO
was running at a speed of 75\,Hz using 18 subapertures and correcting
18 Modes\footnote{We were using so-called ``sensor modes'', the lower
  ones of which are identical to the corresponding Kahunen-Loeve
  modes. The upper sensor modes are linear combinations of higher
  KL-modes to make use of more statistically independent information
  than contained in the original KL-modes\cite{kasper:00s}}.  Albeit, we were far from
the diffraction limit (see Fig.~\ref{fig:ngc6611}). Total integration
times were 5 minutes in each of the filters $J$, $H$ and $K'$. Data
reduction followed standard IR procedures, using a mosaic pattern to
acquire a sky frame and subtracting that before flatfielding. Bad
pixels were mostly removed during mosaic combination, few remaining
bad pixels in non-overlapping regions were removed by filtering.
Photometric calibration was performed by observing the standard
AS37\cite{hunt:98}. For remarks on image quality see the caption of
Fig.\ \ref{fig:ngc6611}

\subsection{Determination of the IMF}
The whole of NGC 6611 was thoroughly examined by Hillenbrand et al.
(1993)\cite{hillenbrand:93}.  They give a mean age for the cluster of
2($\pm 1$)\ttt{6}\,yrs and a mean extinction of 3.2\,mag towards the
cluster region. Relying on their age determination, we can de-redden
the stars via the two-colour and the colour-magnitude diagrams.  Thus
we will be able to extend the IMF towards lower mass stars than
Hillenbrand et al. Photometry was done using the IDL-implementation of
DAOPHOT\cite{stetson:87}.  The extinction towards the cluster can of
course vary across the cluster and especially towards the centre -
which we observed.  Fig.\ \ref{fig:ngc6611} shows a distinct lack of
stars close to the central WFS-star. This indication of a higher
extinction towards the centre than towards the outer parts is also
accompanied by indications for high reddening towards the centre (up
to $A_V=15$\,mag). The process of de-reddening the sources and
constructing the IMF is currently under way, the result will be
published in a forthcoming paper\cite{eisenhau:00}.

\section{Identifying the ionising source of G11.11-0.40}

   \begin{figure}
   \begin{center}
   \begin{tabular}{c}
   \psfig{figure=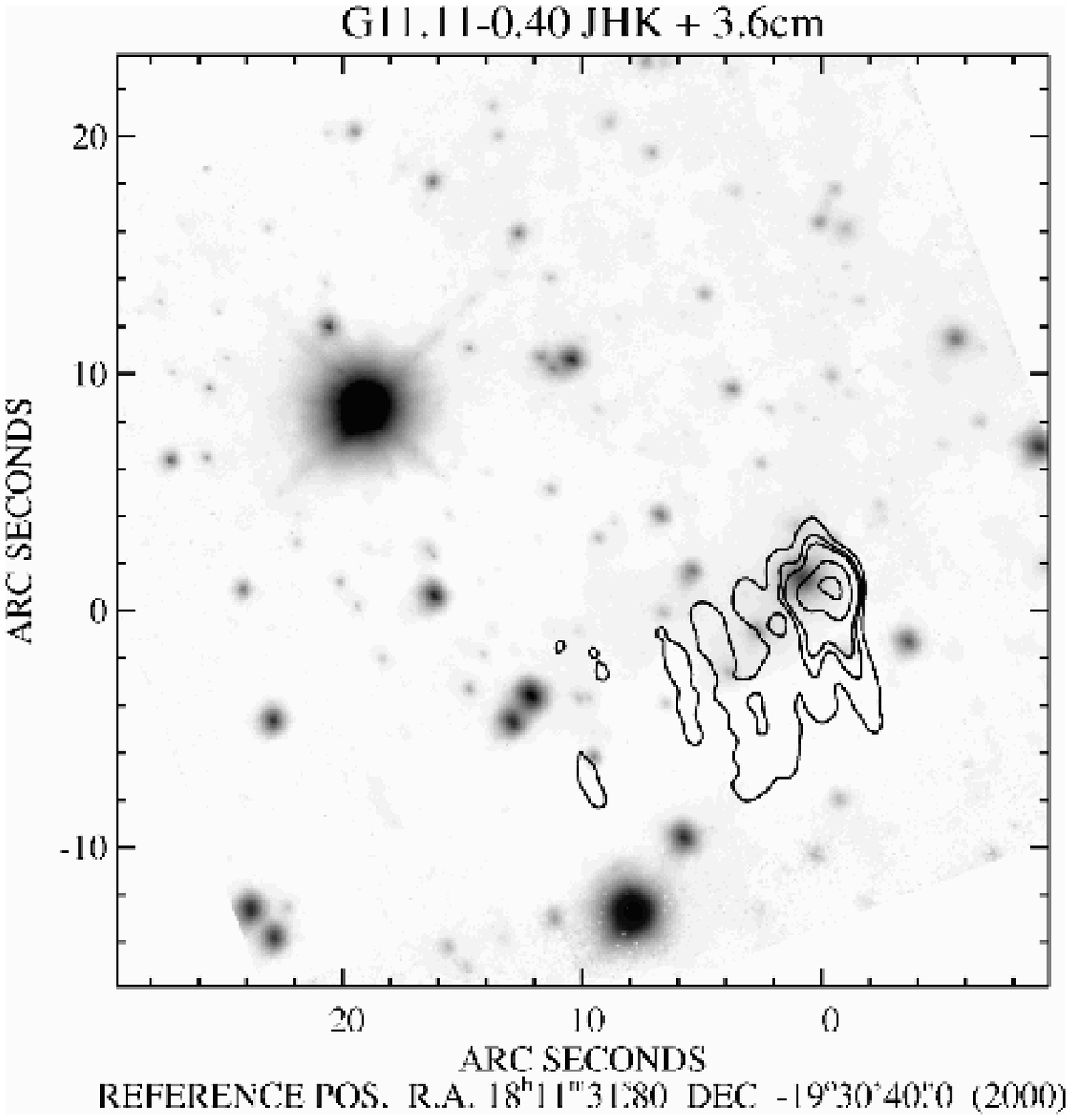,width=16cm} 
   \end{tabular}
   \end{center}
   \caption[G11] 
   { \label{fig:g11} The \uchii G11.11-0.40. Superimposed on the
     $JHK'$ colour composite (The image appears in colour in the
     electronic version only; please check the ALFA web page at
     http://www.mpia-hd.mpg.de/ALFA) are 3.6\, cm VLA contours from
     KCW94.  The achieved resolution on the AO reference star (WFS
     star) is 0\farcs19\,/\, 0\farcs17\,/\,0\farcs24 in
     $K'$\,/\,$H$\,/\,$J$.  This corresponds to Strehl numbers of
     17\%/8\%/3\%.  Due to the limited size of the isoplanatic patch,
     these values are no longer valid at the location of our target.
     Here, at the reference position, we reach
     0\farcs44\,/\,0\farcs52\,/\,0\farcs55 (2\%\,/\,2\%\,/\,0\%). The
     brightness of the WFS star is 10.8\,mag in $V$-band, the seeing
     during the exposure was 0\farcs9}
   \end{figure} 
   
   Ultracompact H{\sc ii} regions (UCH{\sc ii}s) are small (0.15\,pc),
   ionised areas surrounding newly born massive stars (m $\ge$
   8\,\msun). Deeply embedded in the natal molecular clouds of these
   stars, UCH{\sc ii}s are invisible at optical wavelengths but appear
   bright from the near infrared to the radio domain. UCH{\sc ii}s
   appear in many different morphologies and one of the key questions
   is why they are so many of them: Comparing the lifetime which
   theory predicts for the expansion of an ionised shell around a star
   to the lifetime of  O-type stars, one computes a predicted number of
   UCHIIs which is at least a factor of ten lower than the observed
   number. Among the possibilities to answer this question and to
   distinguish between the many models that describe UCHIIs, is the
   clear identification of the ionising sources of these objects.

\subsection{Observations and Data Reduction}
G11.11 was observed during the same night as NGC\,6611.  Although
external conditions were practically the same and the WFS-star 2\,mag
fainter than that of NGC\,6611, the AO system was much better adapted
to the seeing this time: In the image centre, the stellar PSF is
practically diffraction limited in $K'$ and $H$ and Strehl numbers
reach 28\% even after the mosaic combination. Observational parameters
and techniques were also identical to NGC\,6611. Total integration
times were 5 minutes in each band in this case. The resulting $JHK'$
colour composite image can be seen in Fig.\,\ref{fig:g11}

\subsection{The Ionising Source}

\begin{figure}
   \begin{center}
   \begin{tabular}{c}
   \psfig{figure=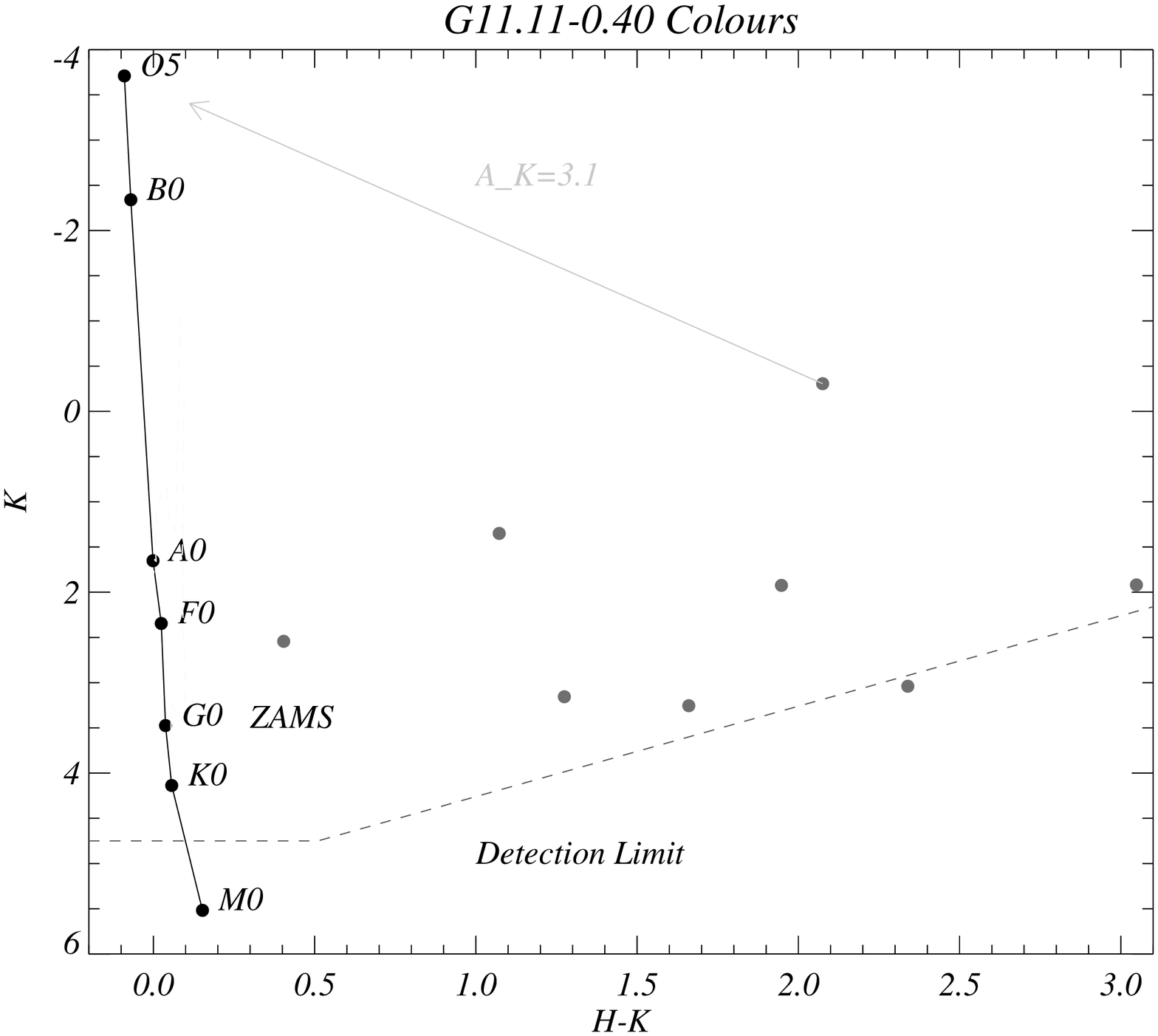,width=12cm,angle=90} 
   \end{tabular}
   \end{center}
   \caption[G11] 
   { \label{fig:g11.cm} Colour-magnitude diagramme of G11.11-0.40. the
     locations of the sources in and close to the VLA map are given by
     the grey circles.  The distance module for G11's distance of
     5.2\,kpc has been applied to the magnitude values. The brightest
     object inside the VLA map has a dereddening arrow.  This arrow
     gives the de-reddening vector for the extinction measured from a
     comparison between Br$\gamma$ and 2\,cm free-free flux.}
   \end{figure}

   Figure \ref{fig:g11} shows that clearly three rather red point
   sources\footnote{At our resolution of 0\farcs4} are located close
   to the ionised region denoted by the contours of the 3.6\,cm
   free-free emission. They appear to lie on the northern rim of a
   dark cloud which clearly blocks background stars to the south. To
   identify the ionizing source of G11, photometry was performed on
   the $H$ and $K´$ frames of G11. the sources inside the ionised
   region (see VLA contours in Fig.  \ref{fig:g11}) are not detected
   in $J$. For photometry, an IDL adaption of DAOPHOT was used. The
   brightest source inside the ionised region was used as PSF
   reference for DAOPHOT.  Since no signs for resolved aditional
   emission was detected from this source, this step seemed justified.
   It also guarantees optimum adaption to the local shape of the PSF,
   which is slightly elongated due to tip-tilt anisoplanatism.  The
   results of the photometry are shown in the colour-magnitude
   diagramme in Fig.\,\ref{fig:g11.cm}. This figure shows the zero age
   main sequence (ZAMS) in red plus some pre-main sequence
   evolutionary tracks by D'Antona \& Mazitelli\cite{dantona:94}.  For
   the sources inside G11, a distance module of 13.58, corresponding
   to a distance of 5.2\,kpc\cite{kurtz:94} was applied to the
   $K'$-magnitude.

\begin{figure}
   \begin{center}
   \begin{tabular}{c}
   \psfig{figure=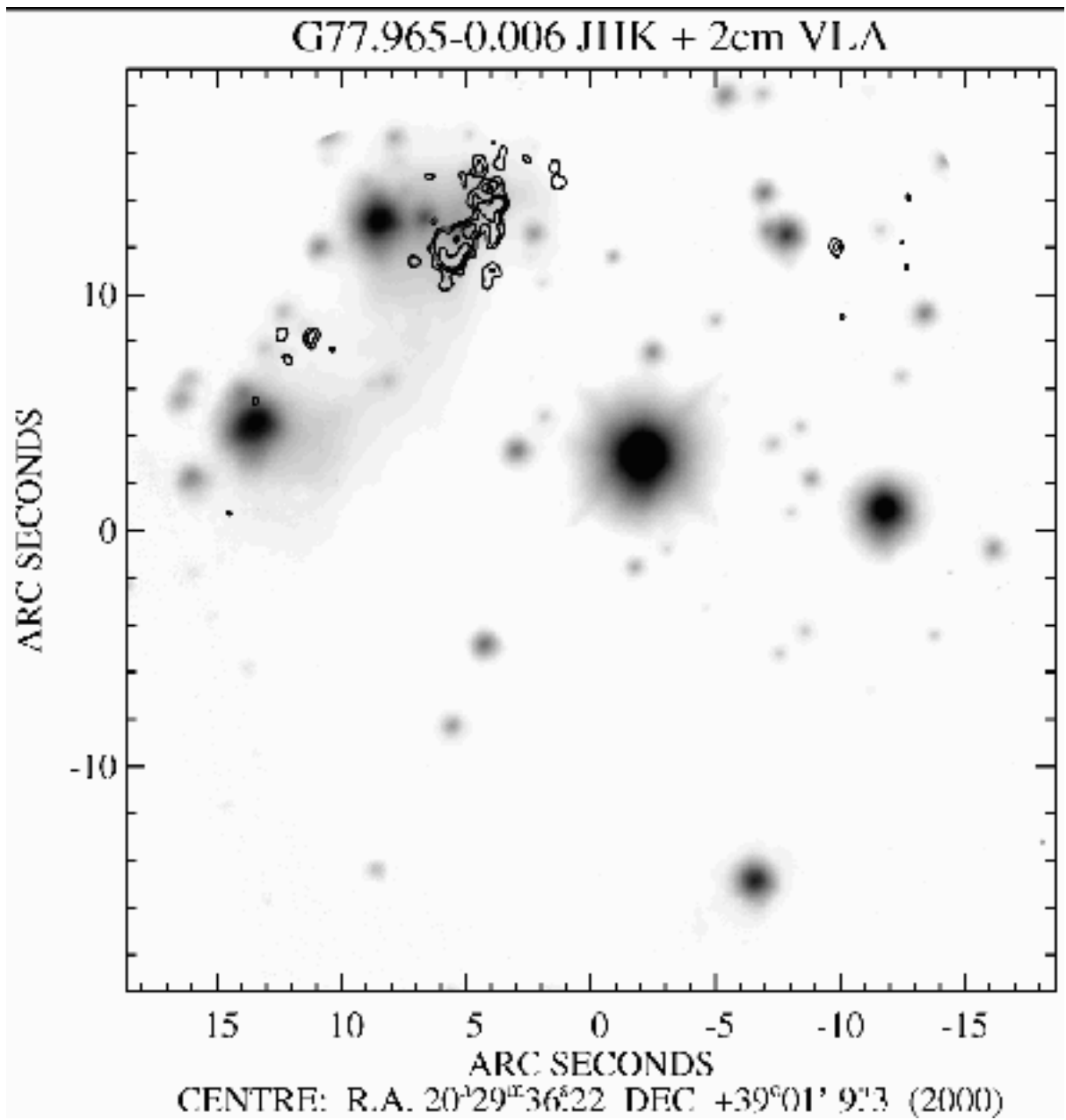,width=16cm} 
   \end{tabular}
   \end{center}
   \caption[G77] {\label{fig:g77} $JHK'$ coulour composite image of the \uchii
     G77.97-0.01 (The image appears in colour in the electronic
     version only). This image shows what ALFA can achieve under
     really limiting conditions: The seeing during the exposure was
     around 1\farcs5 ($H$-band!) and the brightness of the WFS-star is
     only 11.3\,mag ($V$-band). Albeit, the maximum resolutions are
     0\farcs32\,/\,0\farcs52\,/\,0\farcs72 in $J$\,/\,$H$\,/\,$K'$ on
     the WFS star. This corresponds to Strehl numbers of 6\%, 2\%, and
     0\%.  On the target itself, denoted by the superimposed VLA
     contours from KCW94, the Strehl numbers are essentially all 0,
     but the reolutions are still 0\farcs43, 0\farcs66, and 0\farcs92.
     Be reminded, that again all numbers are for the finally reduced
     and combined image.}
   \end{figure}
 
   The 8 sources inside or closest to the ionised region are shown
   here. From a Br$\gamma$ map observed with IRAC2b on ESO's 2.2\,m
   telescope on La Silla (Chile), we derived the extintion towards the
   ionised region via comparison to the observed free-free flux at
   2\,cm. This procedure yielded an extinction of 3.1\,mag at the
   wavelength of Br$\gamma$ (2.166\,\mum). This extinction is assumed
   to be constant across the $K´$-band. When applying the
   corresponding de-reddening vector to the brightest source inside
   G11, we find the star located just short of the position of an O5 star
   in Fig.\,\ref{fig:g11.cm}.
   
   When calculating the Emission measure from the 2\,cm emission, we
   get 0.7\ttt{6}\,pc\,{\rm cm\ensuremath{^{-6}}\xspace},
   corresponding to an electron density of 4.2\ttt{3}\,\cmc. According
   to Kurtz et al.\cite{kurtz:94}, this request a minimum 10$^{47.7}$
   Lyman continuum photons per second from the ionizing source. Models
   by Panagia\cite{panagia:73} indicate that this number can be
   delivered by a single star around spectral type O9. Of course one
   has to be careful:  On the one hand, it is generally unknown how
   much the UV photon flux is influenced by the presumed youth of the
   star. On the other hand, dust inside the region may absorb Lyman
   continuum photons and thus require a star of earlier spectral type
   than O9. Such dust would, if located between the Br$\gamma$
   emitting region and the star itself also explain the additional
   reddening between the de-reddened location of the ionizing source
   in Fig.\,\ref{fig:g11.cm} and the actual location of an O5 star.
   
   Information on the other sources, on dust masses/densities and the
   IMF close to G11 will be given in Henning et al.\cite{henning:00}

\section{Spectroscopy of the young binary system T\,Tau}
\begin{figure}
   \begin{center}
   \begin{tabular}{c}
   \psfig{figure=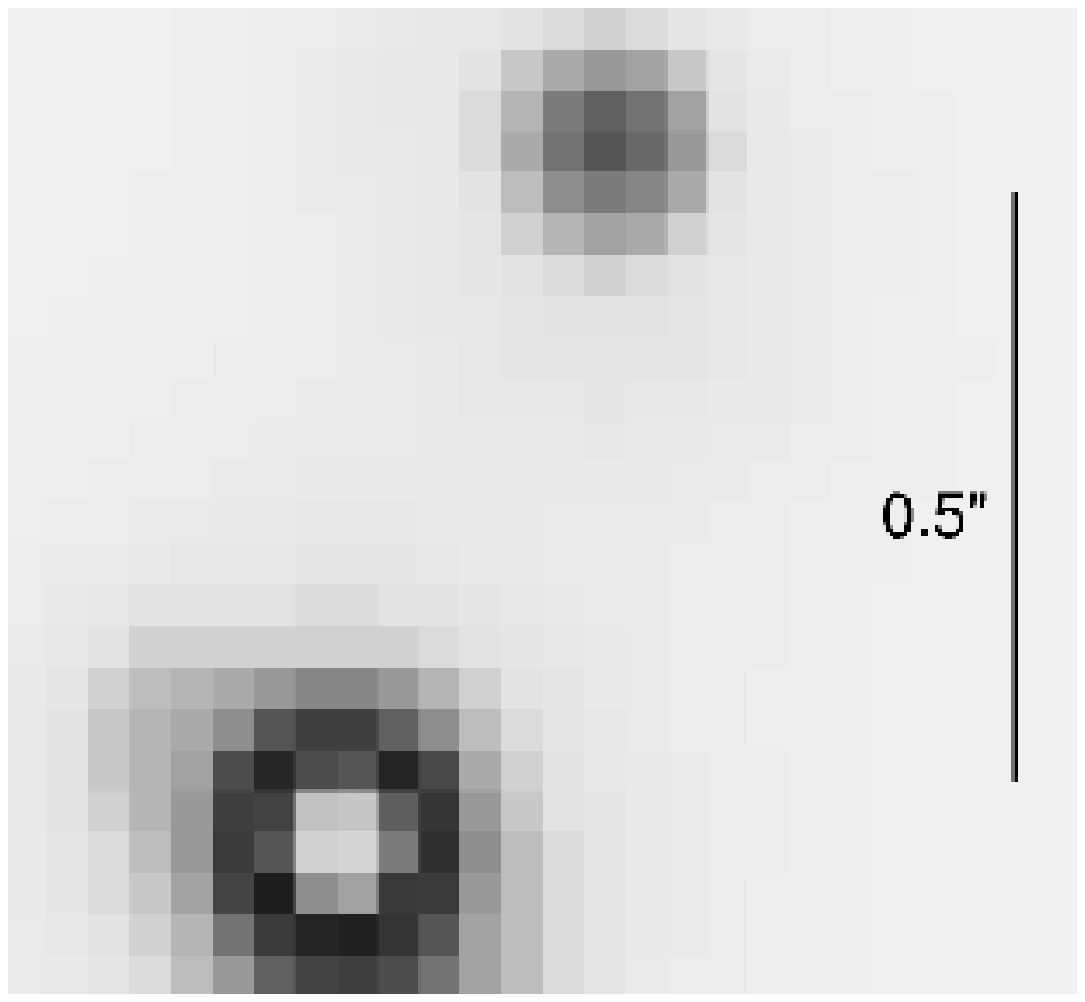,width=9cm} 
   \end{tabular}
   \end{center}
   \caption[TTi] {\label{fig:tti} Reconstructed $K'$-image of T\,Tau taken with the 
     3D and ALFA instruments. The FWHM is 0\farcs14, a Strehl number
     cannot be calculated because the total flux cannot be measured
     from this 1\arcsec$\times$ 1\arcsec field of view. Spectra have
     been extracted on the central 2$\times$ 2 pixels of each PSF. The
     separation of the two components of the T\,Tau system is
     0\farcs69}
   \end{figure} 
   
   The AO system ALFA can be operated in combination with the integral
   field sepctrometer 3D\cite{weitzel:96}. This combination was used
   in September 1999 to observe the young binary system T\,Tau. ALFA
   delivered a resolution of 0\farcs14. Spectra were extracted from
   the central 2$\times$2 pixels of each PSF. The spectral resolution
   of 3D is 1000 in this mode, the spectral range extends across
   0.4\,\mum and thus across the complete $H$ and $K$-bands.

\begin{figure}
   \begin{center}
   \begin{tabular}{c}
   \psfig{figure=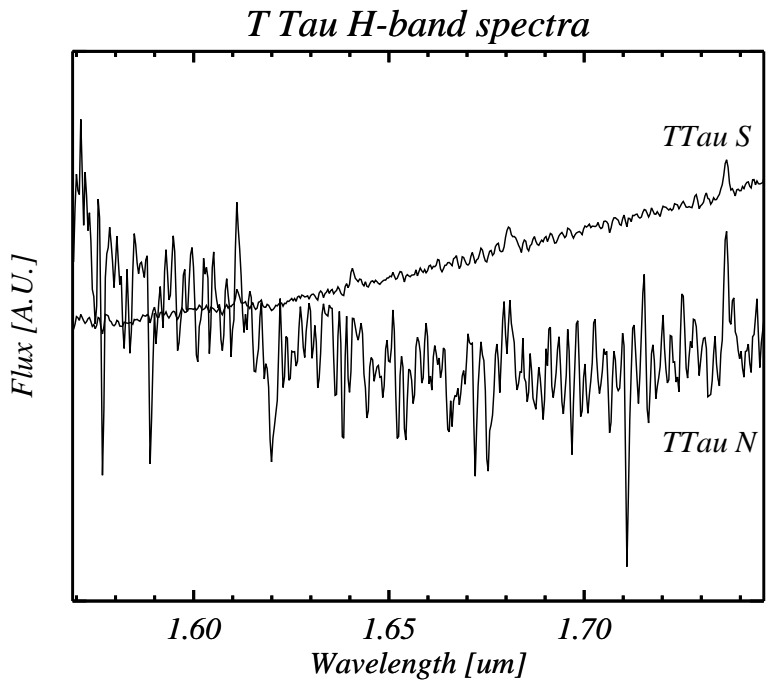,width=12cm,angle=90} 
   \end{tabular}
   \end{center}
   \caption[TTh] {\label{fig:tth} $H$-band spectra of the 
     T\,Tau system.}
   \end{figure} 

\subsection{The Spectra}
\begin{figure}
   \begin{center}
   \begin{tabular}{c}
   \psfig{figure=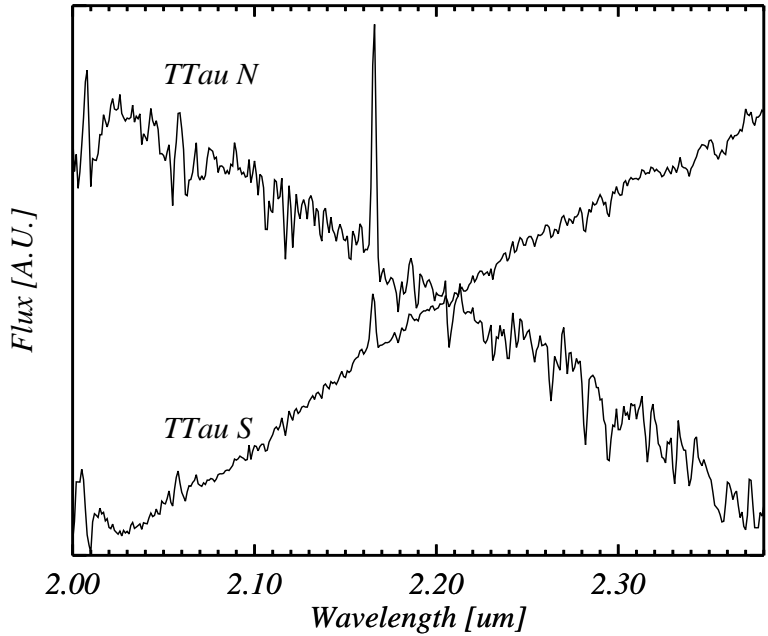,width=12cm,angle=90} 
   \end{tabular}
   \end{center}
   \caption[TTk] {\label{fig:ttk} $K$-band spectra of the T\,Tau system. }
   \end{figure}

A thourough introduction into the T\,Tau system can be found e.g. in \cite{herbst:96}.
Resulting spectra can be seen in figures \ref{fig:tth} and
\ref{fig:ttk}. The primary or northern component (N) has its maximum
shortwards of the $H$-band and shows a falling spectral energy
distribution (SED) across the $K$-band. These are the first clearly
separated spectra of both T\,Tau components. The behaviour of the
secondary or southern component (S) is exactly the opposite. Its SED
is steeply rising across both bands.  Nevertheless, the presence of
pronounced \brg lines indicates, that both objects are actively
acreting, classical T\,Tauri stars.

As for the secondary, line emission from higher Bracket transitions
can be seen in the $H$-band data. These data are currently undergoing
a thorough analysis and will be presented in a forthcoming
paper\cite{kasper:00}

\section{The Impact}

We have presented data obtained during the Science demonstration
programme of the ALFA system in 1999. These deal with various aspects
of star formation research. The analysis of these data currently under
way would have been impossible, had they been taken under seeing
conditions (1\arcsec) at the time of the observations. The data
quality reaches from that of corresponding to data taken at very good
telesopes at very good sites (remember, we only have a 3.5\,m
telescope on an average mountain) to better than HST (See the
spectroscopic data of T\,Tau). Clearly, the picture of star formation
will be complemented by the detailed studies of IMFs in clusters and
UCH{sc ii}s, by the studies of ionisation mechanisms, and by highly
detailed studies of individual systems like T\,Tau. From our results,
it becomes clear that any study relying on observational results of a
quality inferior to ours must lead to misleading or wrong results.
Binary misidentifications, missing photometric completeness due to low
sensitivity when observing with low resolutions, misinterpretations of
sprectal data taken in close environments of stars, all these errors
are inevitable consequences of observing at natural seeing conditions
of 0\farcs7 or worse.

\acknowledgments     
 
The authors would like to thank the rest of the ALFA and 3D teams for
their co-operation during the observing runs.


  \bibliography{/home/feldt/Opus/opus.bib}   
  \bibliographystyle{spiebib}   
 
  \end{document}